\documentclass[10pt,journal,letterpaper,compsoc]{IEEEtran}

\newcommand{\eg}{{\it e.g., }}

\newcommand{\name}{{E2C }}

\usepackage{balance}
\usepackage{times}
\usepackage{cite} 
\usepackage{pifont}
\usepackage{color}
\usepackage{epsfig}
\usepackage{float}
\usepackage{graphicx} 
\usepackage{amsmath}
\usepackage{url}
\usepackage{subcaption}
\usepackage{amsthm}
\usepackage{algorithmicx}
\usepackage{algpseudocode}
\usepackage{algorithm}
\usepackage{xspace}
\usepackage{xcolor}
\usepackage{comment}
\usepackage{bbding}
\usepackage{pifont}
\usepackage{wasysym}
\usepackage{amssymb}
\usepackage{multirow}
\newcommand{\xmark}{\ding{55}}%

\usepackage{array,booktabs}
\usepackage{tabularx}

\usepackage[colorlinks, citecolor=blue]{hyperref}

\begin{document}
\title{E2C: A Visual Simulator to Reinforce Education of Heterogeneous Computing Systems}

\author{Ali Mokhtari, Drake Rawls, Tony Huynh, Jeremiah Green, Mohsen Amini Salehi\\

High Performance Cloud Computing (HPCC) Laboratory,\\ School of Computing and Informatics,\\ University of Louisiana at Lafayette, LA 70503, USA\\
 E-mail: \{ali.mokhtari1, drake.rawls1, tony.huynh1, jeremiah.green1, amini\}@louisiana.edu
    }

\maketitle
\begin{abstract}
 Heterogeneity has been an indispensable aspect of distributed computing throughout the history of these systems. In particular, with the increasing popularity of accelerator technologies (e.g., GPUs and TPUs) and the emergence of domain-specific computing via ASICs and FPGA, the matter of heterogeneity and understanding its ramifications on the system performance has become more critical than ever before. However, it is challenging to effectively educate students about the potential impacts of heterogeneity on: (a) the performance of distributed systems; and (b) the logic of resource allocation methods to efficiently utilize the resources. Making use of the real infrastructure (such as those offered by the public cloud providers) for benchmarking the performance of heterogeneous machines, for different applications, with respect to different objectives, and under various workload intensities is cost- and time-prohibitive. Moreover, not all students (globally and nationally) have access or can afford such real infrastructure. To reinforce the quality of learning about various dimensions of heterogeneity, and to decrease the widening gap in education, we develop an open-source simulation tool, called \name, that can help students researchers and practitioners to study any type of heterogeneous (or homogeneous) computing system and measure its performance under various system configurations. To make the learning curve shallow, E2C is equipped with an intuitive graphical user interface (GUI) that enables its users to easily examine system-level solutions (scheduling, load balancing, scalability, etc.) in a controlled environment within a short time and at no cost. In particular, E2C is a discrete event simulator that offers the following features: (i) simulating a heterogeneous computing system; (ii) implementing a newly developed scheduling method and plugging it into the system, (iii) measuring energy consumption and other output-related metrics; and (iv) powerful visual aspects to ease the learning curve for students. We used E2C as an assignment in the Distributed and Cloud Computing course. Our anonymous survey study indicates that students rated \name with the score of 8.7 out of 10 for its usefulness in understanding the concepts of scheduling in heterogeneous computing. Moreover, 
 our pre- and post-evaluations indicate that \name has improved the students' understanding of heterogeneous computing systems by around 18\%.
\end{abstract}

\section{Introduction}
Heterogeneity has been an indispensable aspect of distributed computing throughout the history of these systems. In the modern era, as Moore's law is losing momentum due to the power density and heat dissipation limitations~\cite{taylor2012dark, esmaeilzadeh2011dark}, heterogeneous computing systems have attracted even more attention to overcome the slowdown in Moore's law and fulfilling the desire for higher performance in various types of distributed computing systems. In particular, with the increasing prevalence of accelerator technologies (\eg GPUs and TPUs) and the emergence of domain-specific computing via ASICs \cite{taylor2020asic} and FPGA \cite{bobda2022future}, the matter of heterogeneity and harnessing it has become a more critical challenge than ever before to deal with. 

Examples of heterogeneity can be found in any type of distributed system. Public cloud providers offer and operate based on a wide variety of machine types. Hyperscalers such as AWS and Microsoft Azure provide computing services ranging from general-purpose X86-based and ARM-based machines to FPGAs and accelerators~\cite{Amazon_SageMaker}. In the context of Edge computing, domain-specific accelerators (ASICs and FPGA) and general-purpose processors are commonly used together to perform near-data processing, thereby, unlocking various real-time use cases (\eg edge AI and AR/VR applications)~\cite{google_glass, Qualcomm}. In the HPC context, deploying various machine types with different architectures on HPC boards to fulfill the power and performance requirements is becoming a trend~\cite{cardwell2020truly}.

Heterogeneity plays a key role in improving various performance objectives of distributed systems, such as cost, energy consumption, and QoS. That is why harnessing system heterogeneity has been a longstanding challenge in distributed systems (\eg \cite{mokhtari2020autonomous,DENNINNART202046,ipdps19}), and educating it to Computer Science/Engineering (and more broadly STEM) students, and researchers has become necessary. Making use of real infrastructure (such as those offered by the public cloud providers) for benchmarking the performance of heterogeneous systems, for different applications, with respect to different objectives, and under various workload intensities is cost- and time-prohibitive. As an example, consider an IoT-based system that offers multiple smart applications to its users (\eg object detection, face recognition, speech recognition, etc.); there exists a wide range of machine types with different architectures (such as x-86 or ARM-based multi-core CPUs, different types of GPUs, FPGAs, and ASICs) that can process these services. To find an optimal configuration, a student must examine all permutations of these configurations. Moreover, there can be multiple workload intensities and scheduling policies that can affect performance of the system and the student must examine them too. Last but not least, learning about the energy consumption of the heterogeneous computing system in question adds another dimension to the evaluation process that needs to be conducted by the student. 

To avoid the burden of examining all cases, we need simulation tools that can help the students and researchers to study the performance of various system configurations and effectively learn about impacts of heterogeneity in a distributed system. To that end, in this paper, we introduce \name that is an open-source discrete event simulator that simulate any type of heterogeneous (and homogeneous) computing system. By using \name, the students can easily examine their system-level solutions (scheduling, load balancing, scalability, etc.) in a controlled environment within a short time and at no cost. In particular, \name offers the following features: (i) defining user-defined workload generation scenarios with various number of applications (a.k.a. task types) and arrival intensities; (ii) simulating a heterogeneous computing system; (iii) implementing a newly developed scheduling method and plugging it into the system, (iv) measuring power and other output-related things, and (v) visual aspects to ease the learning curve for students.
These features help students who study resource allocation solutions in distributed systems to test and evaluate their solutions easier and faster. Moreover, the graphical user interface would help students to gain a deeper knowledge of resource allocation procedures in distributed computing systems. 

We used \name as an assignment in our Distributed and Cloud Computing class to examine various types of scheduling methods for heterogeneous (and homogeneous) systems under various workload intensities. We conducted a survey on the learning outcomes of the simulator and its usability aspects. Analysis of the survey results showed that the students on average rated \name with the score of 8.7 out of 10 for its usefulness in comprehending scheduling methods for heterogeneous and homogeneous computing systems under different workload intensities. Moreover, based on the survey results, students assessed that \name is easy to use with the average score of 8.3 out of 10, and they evaluated their willingness for recommending \name to others with the average score of 8.3 out of 10.

In the rest of this paper, in Section~\ref{sec:positioning}, we first position \name with respect to other existing simulators. Next, we elaborate on the features of \name in more detail in Section~\ref{sec:sim}. Then, in Section~\ref{sec:assign}, we describe our experience of using \name as a class assignment for Computer Science and Engineering students. In Section~\ref{sec:eval}, the evaluation of \name and the results we obtained are discussed. Availability of the simulator and the conclusion and future extensions of \name are explained in Sections~\ref{sec:code} and \ref{sec:conc}, respectively.

\section{Positioning \name with respect to other existing simulators}\label{sec:positioning}
There are several existing cloud simulators that have been developed to provide researchers and developers with a platform to simulate and test cloud computing environments. Some of the popular cloud simulators include CloudSim~\cite{calheiros2011cloudsim}, EdgeCloudSim~\cite{sonmez2018edgecloudsim}, iFogSim~\cite{gupta2017ifogsim}, iCanCloud~\cite{nunez2012icancloud}, and TeachCloud~\cite{jararweh2013teachcloud}. Table~\ref{tbl:compare} provides a quick positioning of \name with respect to these simulation tools. 


\begin{table} [t]
    \centering

\begin{tabular}{  p{0.1\textwidth} || p{0.05\textwidth}| p{0.03\textwidth}| p{0.08\textwidth}| p{0.08\textwidth} p{0.0006\textwidth}}
\hline

 \textbf{Simulator} &  \textbf{Prog. Language} &  \textbf{GUI} &   \textbf{supporting heterogeneous computing} &  \centering \textbf{workload generator} & \\ \hline

 CloudSim & \centering Java & \centering \xmark & \centering \xmark & \centering limited & \\ \hline
 iFogSim & \centering Java & \centering \xmark & \centering \xmark& \centering limited &  \\ \hline
 EdgeCloudSim & \centering Java & \centering \xmark&\centering \xmark&\centering \checkmark & \\ \hline
 iCanCloud & \centering C++ & \centering \checkmark&\centering \xmark&\centering \xmark  & \\ \hline
TeachCloud & \centering Java & \centering \checkmark&\centering \xmark&\centering limited & \\ \hline
 \hline
E2C & \centering  Python & \centering \checkmark&\centering \checkmark&\centering \checkmark & \\
\hline 
\end{tabular}
\caption{Positioning of \name with respect to other simulation tools for distributed systems. }\label{tbl:compare}
\end{table}

CloudSim is a popular open-source framework used for modeling and simulating cloud computing environments and applications. With its modular and extensible architecture, CloudSim allows users to customize and configure different aspects of the simulation, such as virtual machine management, workload scheduling, and resource allocation policies, to suit their research needs. However, as a Java-based framework, it needs the user-input in the form of Java lines of the code within the back-end for configuring and customizing the environment. As a results, the users (e.g. students) should already have background experience and knowledge of Java and object-oriented programming (OOP). While this can provide its own kind of learning experience, it is not necessarily helpful for teaching about cloud computing and distributed systems, and may only slow down education concentrated in that area. EdgeCloudSim is another simulation platform that is tailored to Edge computing systems. EdgeCloudSim is based on the CloudSim with more functionalities in network modeling and load generator. iFogSim is another CloudSim-based simulation tool that is utilized for modelling and simulation of Fog computing environments, and evaluating the efficiency of different resource management policies in terms of latency (timeliness), energy consumption, network congestion and operational costs. iCanCloud is a cloud computing simulation framework for generating and customizing a large distributed computing system written in C++. iCanCloud comes with a user-friendly GUI which is useful in managing pre-configured virtual cloud systems and generating graphical reports. Although iCanCloud supports consistent heterogeneity in terms of configuring VMs with varying number of CPU cores, it does not support inconsistent heterogeneity by having VMs with accelerators (e.g. GPUs and FPGAs). Jararweh et al. developed a simulation toolkit, called TeachCloud~\cite{jararweh2013teachcloud}, for cloud computing environment equipped with a GUI that allow students easily create the main components in the cloud system. However, TeachCloud lacks supporting the heterogeneous computing systems.
\begin{figure*}[ht!]
    \centering
    \includegraphics[ height=0.55\textwidth]{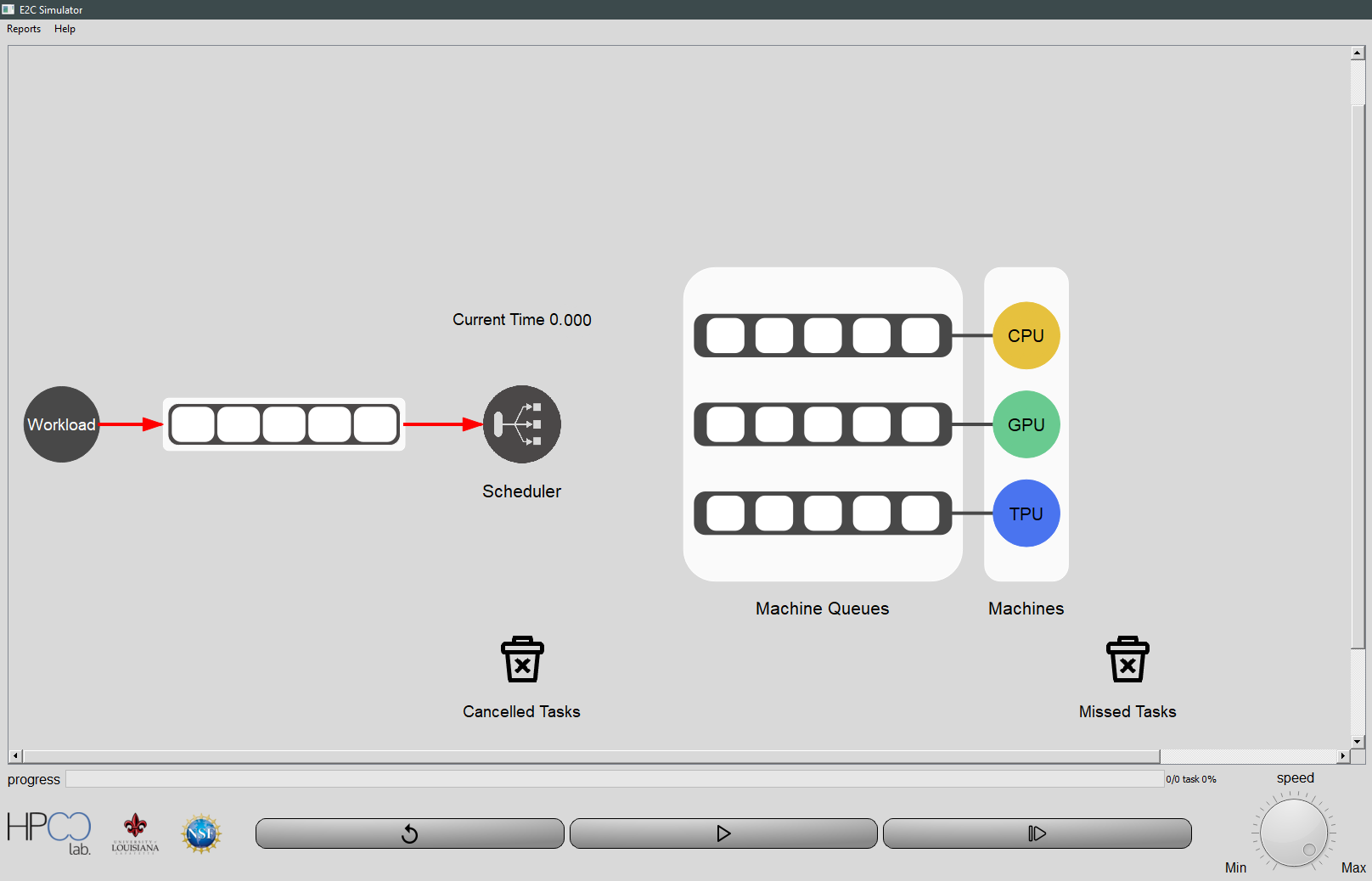}
    \caption{Overview of the \name Simulator that includes major components, being the source workload, a batch queue of arriving tasks, scheduler (a.k.a. load balancer), and a set of heterogeneous machines, represented with different colors. Each machine has a ``machine queue'' where the assigned tasks are queued for the execution.}
    \label{fig:overviewl}
\end{figure*}
In general, these simulators provide valuable insights into the performance and efficiency of cloud computing systems, enabling researchers and developers to make informed decisions when designing and implementing cloud-based solutions. However, there are limitations in using these simulators as an educational tool for teaching heterogeneous distributed computing systems. As per limitations of existing simulators, they lack either a user-friendly graphical interface to make use of the simulator easy and intuitive for the students or supporting heterogeneous computing systems. To overcome these limitations, we developed \name, a simulator explicitly designed for heterogeneous computing systems with an intuitive graphical user interface (GUI). \name is intended to facilitate the study of heterogeneous computing systems for students by enabling them to simulate and explore the characteristics of this type of computing systems through a user-friendly interface.

\name comes with a GUI, that requires no programming input from the user. All inputs can be done directly from the GUI. In addition, the GUI displays simulations in live time, making it well suited for education, as many students perform better through visual learning. \name also aims to be granular, allowing the user to configure the system in many specific ways. This is also important for researchers, given that a simulated system should be highly configurable in order to handle a wide variety of workloads.

\section{Simulating a Heterogeneous Computing System via \name}\label{sec:sim}
Figure~\ref{fig:overviewl} shows an overview of the \name simulator that includes the following major components: (i) workload, (ii) batch queue, (iii) scheduler, (iv) machine queue, and (v) a set of (homogeneous or heterogeneous) machines. In addition, there are two more components that contain canceled and dropped tasks. This is to support circumstances where tasks have hard deadlines and there is no value in executing them beyond their deadline.


A workload is defined as a large group of tasks where each task is a request for an application (task type). In the real world, a heterogeneous computing system can be configured to execute several task types. For instance, a heterogeneous system processing satellite images should support task types for object detection, noise removal, and image enhancements to be performed on the received images. \name enables us to define the task types, arrival distribution for each task type, and their arrival duration. Each task in the generated workload of \name has an arrival time and deadline as well. 

\begin{figure}
    \centering
    \includegraphics[width=0.40\textwidth]{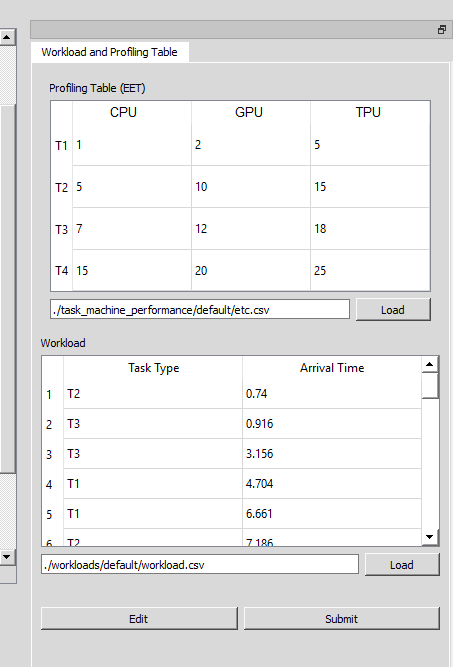}
    \caption{Workload component. Here, the user can load EET and Workload CSV files. The user can also modify the EET matrix and arrival times of the task with the ``Edit'' button. Upon loading new CSV files or editing values, the user must press the ``Submit'' button. EET and Workload files must be compatible. \texttt{T1, T2, T3} represent different task types in this simulation.}
    \label{fig:workload}
\end{figure}

The machines in the distributed system can be identical (homogeneous) or non-identical (heterogeneous).
Note that the heterogeneity of the system is modeled by a matrix, called the Expected Execution Time (EET) matrix \cite{ali2000representing, panda2019energy, panda2015efficient}. This matrix defines the expected execution time of each task type on each machine. This is to model a real world heterogeneous system, where any given task type (e.g., object detection, noise removal, etc.) is expected to have a differing execution time across heterogeneous machines. The opposite holds true for a homogeneous system where any given task type has identical execution time across all machines. As shown in Figure~\ref{fig:workload}, the user has access to the EET matrix by selecting the workload component. Users can either modify the EET matrix manually or load the desired one as a CSV file. 

As shown in Figure~\ref{fig:workload}, the user can load the desired workload trace as a CSV file in this section. The user must keep in mind that the workload trace must conform to the EET matrix. That is, there can be no task type within the workload that is not defined within the EET. Upon the arrival of a task, the simulator transfers the task to the batch queue. The batch queue is where tasks are held before being scheduled. Next, based on the selected scheduling method, the scheduler selects a task from the arrival queue. 

\begin{figure}
    \centering
    \includegraphics[width=0.45\textwidth]{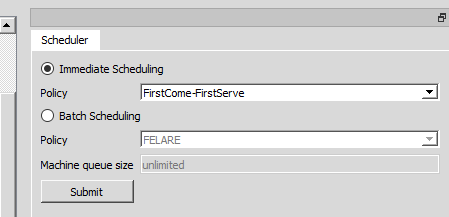}
    \caption{Scheduler component. Here the user may select between the various immediate or batch scheduling policies, along with setting the machine queue size. The machine queue size is limited to infinite for immediate policies, but can be changed for batch policies.}
    \label{fig:scheduler}
\end{figure}

Figure~\ref{fig:scheduler} shows the scheduler options. The user can choose between immediate scheduling or batch scheduling \cite{maheswaran1999dynamic}. Immediate scheduling is when incoming tasks are immediately scheduled to a machine upon arrival, whereas, with batch scheduling, tasks are buffered in the batch queue so the scheduler can make a more informed decision. Typically, immediate mode scheduling methods impose a lower overhead and generally load balancers use this type of scheduling \cite{maheswaran1999dynamic}. The following immediate policies are currently implemented into \name as options: FirstCome-FirstServe (FCFS), Min-Expected-Completion-Time (MECT), and Min-Expected-Execution-Time (MEET). For batch policies, \name currently implements: ELARE, FELARE, MinCompletion-MinCompletion (MM), MinCompletion-MaxUrgency (MMU), and MinCompletion-SoonestDeadline (MSD). An explanation of these methods can be found in \cite{mokhtari2020autonomous}.

\begin{figure}
    \centering
    \includegraphics[width=0.43\textwidth]{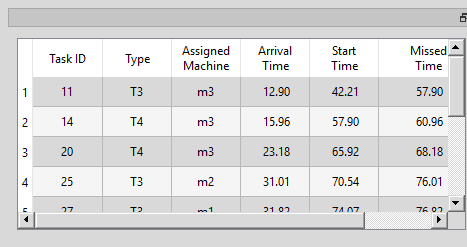}
    \caption{Missed Tasks component shows the task ID that missed its deadline, along with its task type, assigned machine, arrival time, start time, and the time when it missed. }
    \label{fig:missed}
\end{figure}

There exist two options for the scheduled tasks: (i) it might be canceled because of missing its deadline before assignment; or (ii) it might be mapped to one of the available machines. The status of a canceled task is set to ``canceled'' and no more process is needed. The canceled tasks component shows the number of tasks have been canceled so far. In the case of mapping decisions, the task is appended to the local queue of the assigned machine until the machine queue is saturated. Tasks are executed on the assigned machine in a sequential manner by default. If a task missed its deadline while executing on the machine, it is dropped from the machine. As shown in Figure~\ref{fig:missed}, the Missed Tasks component shows the tasks that missed their deadline. 

Importantly, \name is designed to be modular, hence, providing the ability for the user to modify the existing scheduling methods or adding their own custom-designed scheduling methods. This feature is particularly helpful for researchers to examine new methods under various conditions and configurations.

After the user selects and submits the EET and workload, they will press the ``Play" button near the bottom-middle of the GUI. This will begin the animation of tasks flowing from the incoming workload to scheduler to machines, along with the ``Current Time" which will update continuously during simulation. If you press the ``Play" button again during the simulation run-time, the simulation will be paused. The button right of the play button is the ``Increment" button, which when pressed while the simulation is paused will perform the next individual step that would performed (i.e. a task being submitted to a machine by the scheduler, or a task's execution being completed by a machine, etc.). This can be helpful if you wish to analyze each specific action of the simulation. To the left of the ``Play" button is the ``Reset" button, which can be used either during a pause or after completion of a simulation. This will allow you to begin a new simulation, also allowing you load in a new EET and/or workload should you choose. Along with these three options, during the simulation run-time, you can choose to alter the speed at which the simulation runs by using the speed dial located at the bottom right. This can be useful for either getting quicker results or for better visibility of the animated simulation.

Upon completion of a simulation within E2C, the user may view a report, and optionally, save the report as a CSV file. There is an option for a ``Full Report," ``Task Report," ``Machine Report," and ``Summary Report." The Full Report displays the majority of relevant information regarding the simulation - this is the option to view all data related to each task and and how each machine performed on it. The Task Report displays information that is more centric to the individual tasks of the workload, whereas the Machine Report displays data more relevant to the machines of the system. Lastly, the Summary Report displays a summary of the workload data without the specifics of each individual task.



The \name simulator can be implemented as a learning tool for undergraduate and graduate students, and also serve practical solutions for researchers and practitioners. Through E2C, students can gain the ability to analyze, design, implement, and test distributed computer systems and components. They can deeply investigate scheduling methods, how they work, and gain insights into their advantages and disadvantages. Along with this, they can develop their own scheduling method(s) and use \name as a means to implement it. Students can also learn how heterogeneity can improve the performance of the system through defining machines that have better performance for executing specific task types. Moreover, they can study the energy consumption of the system once a certain scheduling method is applied, allowing them to learn about resource management. So far, we have used the \name simulator for students in ``Distributed and Cloud Computing'' courses to examine the impact of different scheduling policies on homogeneous and heterogeneous systems with various workload intensities. Similarly, the simulator can be used for the ``Operating Systems'' and ``Computer Networks'' courses at the undergraduate and graduate levels to teach students about the impact of scheduling at different levels. 


Researchers in the resource allocation area and cloud solution architects can employ the \name simulator to test their solution prior to implementation. Being highly customizable, they can configure E2C to represent its real world counterpart. Through this, they may test the outcome of a heterogeneous system with different scheduling methods without spending real resources, saving both money and time. The outcome to be tested can be things such as QoS through task completion percentage (versus missed and cancelled tasks), energy consumption of machines (resource management), and how different scheduling methods perform on any given system. This way, researchers can apply practical use of \name in order to help design and compare their own real world distributed systems or clusters. As an example, in \cite{ali22}, we have used \name to examine energy efficiency and fairness of scheduling methods on a heterogeneous edge. Also, in \cite{zobaed22}, we extended \name to simulate the memory allocation policies of multi-tenant applications on a homogeneous edge computing system.

\section{Class Assignment for Computer Science and Engineering Students}\label{sec:assign}
The \name was used, and will continue to be used, by undergraduate and graduate students of the University of Lafayette's Distributed Cloud Computing course. Before \name was implemented as an assignment for the students, there were no assignments for evaluating the students' understanding of heterogeneous systems and scheduling methods through simulation. Now, with the addition of E2C, students have a means to learn these subjects through coursework. In this assignment, \name was used to teach students about the impact of various scheduling methods in heterogeneous and homogeneous computing systems operating under various workload intensities. It also asked the graduate students to develop and implement their own scheduling policies and compare it with the existing solutions. The installation and graphical user-interface of \name is user friendly and works on any operating system, which makes it easy to pick up and use for projects or assignments. We have created a web-based documentation\footnote{\smallskip\noindent
\footnotesize \name documentation can be accessed at: \texttt{\url{https://hpcclab.github.io/E2C-Sim-docs/}}
\normalsize} where all the features of the simulator---from installation to reporting---are explained. 

In this assignment, students were to read and learn about the basic components of E2C, being task types, machines, EET matrix, workload trace, and task deadlines. The students would then use the simulator to evaluate the different scheduling methods currently implemented by \name on both a homogeneous and a heterogeneous system. For the homogeneous system, students were to use three workload traces with arrival intensities ranging from \texttt{low, medium,} to \texttt{high} to stress the system at different levels.  For each arrival intensity level, they ran the simulation and saved the CSV output files, provided by \name, summarizing all the data related to the simulation for three different immediate scheduling methods, namely FirstCome-FirstServe (FCFS), Minimum-Expected-Completion-Time (MECT), and Minimum-Expected-Execution-Time (MEET). Students then created a bar graphs to depict the percentage of completed tasks that each scheduling method results under each intensity level. The expected results is that higher intensity workloads lead to a lower completion rate (i.e., more tasks missing their deadlines). In addition to observing this behavior, the students had to analyze and report the behavior of different scheduling methods.
\begin{figure}[h]
    \centering
    \includegraphics[width=0.45\textwidth]{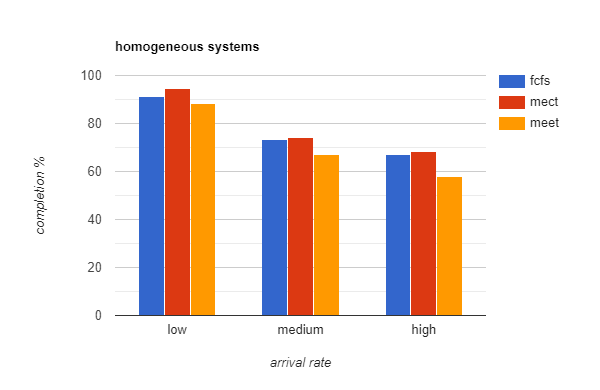}
    \caption{A bar graph with completion \% for immediate scheduling methods on a homogeneous system, showing results for varying intensities using FCFS, MECT, and MEET policies. }
    \label{fig:homo_graph}
\end{figure}

\begin{figure}[h]
    \centering
    \includegraphics[width=0.45\textwidth]{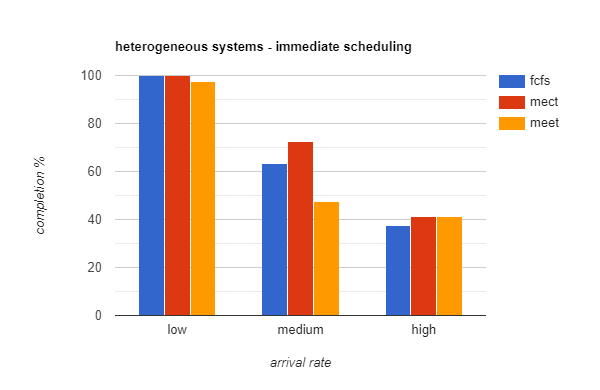}
    \caption{A bar graph with completion \% for immediate scheduling methods on a heterogeneous system, showing results for varying intensities using FCFS, MECT, and MEET policies. }
    \label{fig:het1_graph}
\end{figure}

\begin{figure}[h]
    \centering
    \includegraphics[width=0.45\textwidth]{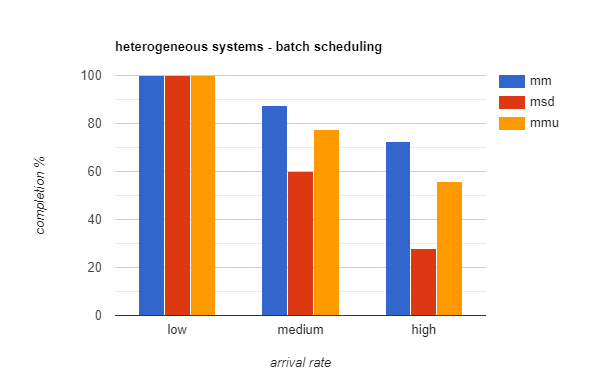}
    \caption{A bar graph with completion \% for batch scheduling methods on a heterogeneous system, showing results for varying intensities using MMU, MSD, and MMU policies. }
    \label{fig:het2_graph}
\end{figure}

\begin{figure*}[ht!]
  \begin{subfigure}{0.5\textwidth}
     \centering
     \includegraphics[width=\textwidth]{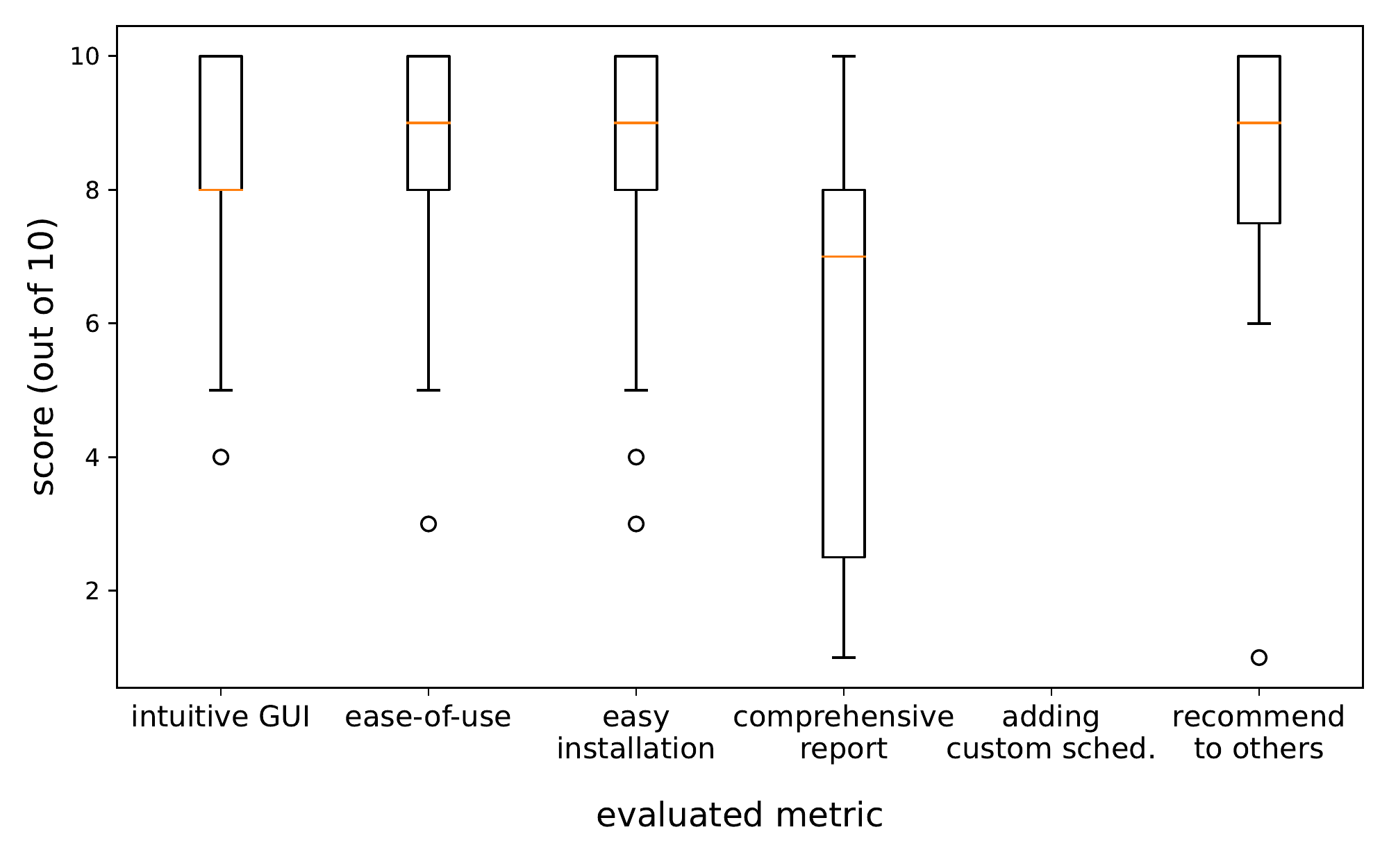}
     \caption{Evaluation of user experience with \name}
     \label{fig:hci}
      
  \end{subfigure}\hfill
  \begin{subfigure}{0.5\textwidth}
     \centering

     \includegraphics[width=\textwidth]{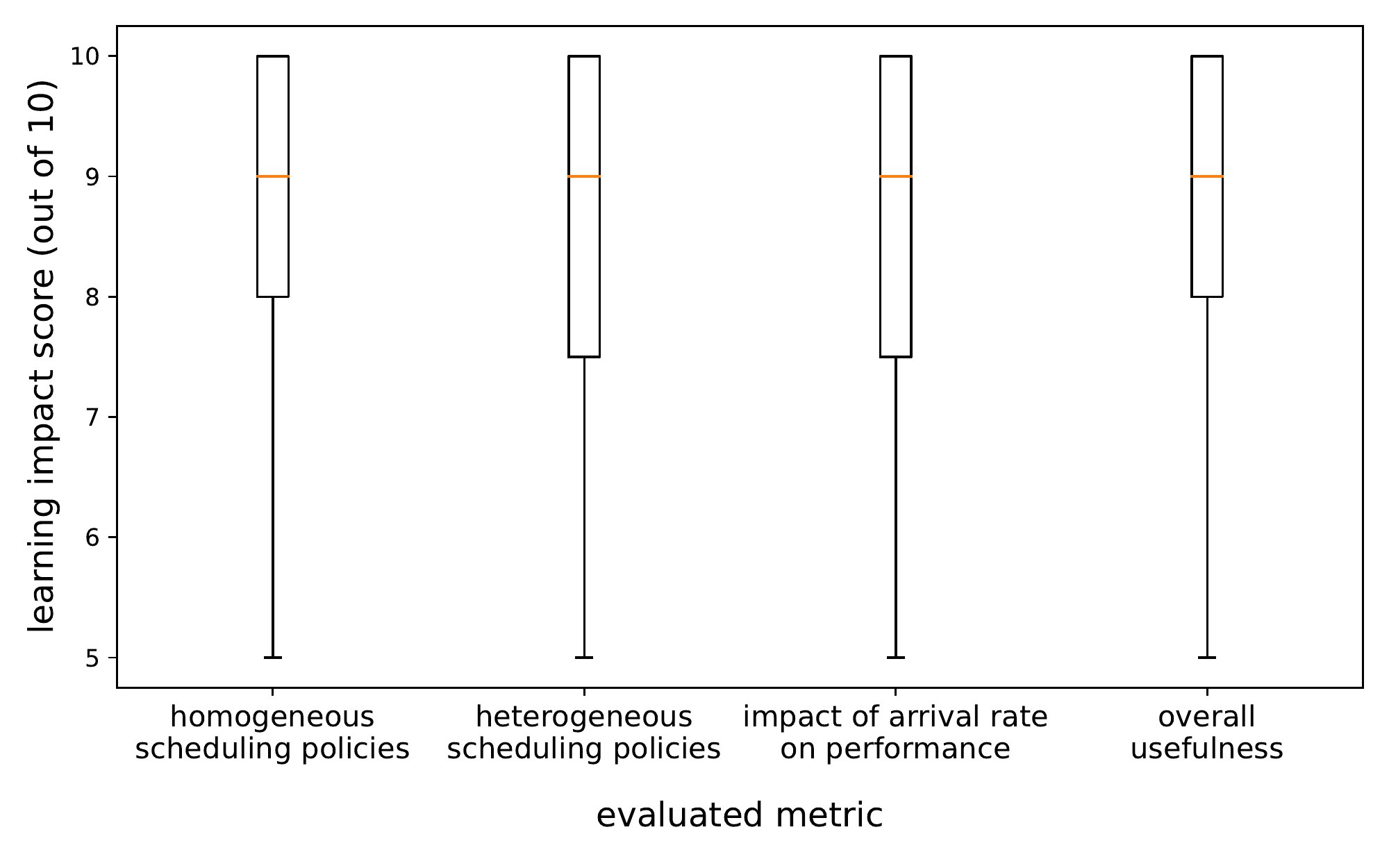}
     \caption{Evaluation of learning objectives  }
     
     \label{fig:learning}
  \end{subfigure}\hfill
  \caption{Illustration of the survey on the students' experience in accomplishing their distributed systems assignment via E2C (a) This subfigure demonstrate the HCI experience of the students with E2C (b) This subfigure shows how much E2C simulator could help students in understanding the characteristics of task scheduling policies in the homogeneous and heterogeneous configurations.}
  \vspace{-3mm}
  \label{fig:survey}
      
\end{figure*}

For the next part of the assignment, they would do similarly but with a 
heterogeneous system instead. For this part, in addition to the immediate scheduling policies, they would also be testing the batch mode policies: MinCompletion-MinCompletion (MM), MinCompletion-MaxUrgency (MMU), and MinCompletion-SoonestDeadline (MSD). Required by the graduate students and optional to undergraduates as a bonus, the third part of this assignment was to create and implement their own scheduling method for the heterogeneous system that enabled fairness across various task types in the system. After these simulations and implementations were complete, students were to perform an analysis of their findings on both the homogeneous system and heterogeneous system, and answer questions that show what they have learned about scheduling and its related methods.

The creation of graphs to evaluate their findings is straightforward due to the way saving data from simulations is within \name. Once a simulation is complete, all students needed to do is go to the reports menu and save the report as a CSV file.

For the bar graphs that the students create for both their findings on homogeneous and heterogeneous systems, they plot the completion percentage (completed tasks/total tasks in workload) for each scheduling method. Some examples of their findings show a bar graph depicting completion percentage for immediate scheduling policies on a homogeneous system (Figure~\ref{fig:homo_graph}), immediate scheduling policies on a heterogeneous system (Figure~\ref{fig:het1_graph}), and batch scheduling policies on a heterogeneous system (Figure~\ref{fig:het2_graph}).


The learning outcomes of this assignment was to understand the impact of different scheduling methods in face of homogeneous and heterogeneous systems, and to analyze the advantages and disadvantages of each. For instance, they analyzed why Minimum-Expected-Completion-Time (MECT) performs better than FirstCome-FirstServe (FCFS) method, and why the batch policies outperform immediate scheduling policies for heterogeneous systems.


\section{evaluating learning outcomes of \name}\label{sec:eval}
As mentioned in Section~\ref{sec:assign}, \name have been examined as an assignment in the Distributed and Cloud computing course. After the assignment, we conducted a survey across the students to evaluate the impact of \name on their learning. 23 students (14 undergraduate students and 9 graduate students) participated in this survey study. The demography of the 23 students are as follows: (i) Gender: 73.9\% students were male and 26.1\% of them were female; (ii) Degree level: 60.9\% students enrolled in Bachelor's degree (undergraduate) and 39.1\% were pursuing higher level of education including master and doctoral degree (graduate); (iii) Programming experience: 
The mean and median values of the students' programming experience are 3.8 and 3 years, respectively; and (iv) Passed Operating System (OS) course: 43.5\% of the students have already completed the OS course and 56.5\% of them have not previously passed that course. The questions of the survey\footnote{\footnotesize{The complete survey can be retrieved from \hyperlink{https://drive.google.com/file/d/1iW3pHFb7Uic-nmlUf6xIA_SZIXf7PoD4/view?usp=sharing}{here}.}} were in two categories: (i) Those related to the user interactions (experience) with the \name simulator that is shown in Figure~\ref{fig:hci}; and (ii) Those focuses on the specific learning outcomes, i.e., how much the knowledge of students was improved as a result of doing this assignment. The result of this category is shown in Figure~\ref{fig:learning}. All students were asked to rate \name with respect to each evaluation metric in the scale of 10. 


The user experience part studies the user-friendliness of the \name interface and how it makes technical concepts intuitive. Installing E2C is the first experience of such nature. Figure~\ref{fig:hci} shows that students on average evaluated that the installation part is an easy and straightforward procedure with the score of 8.3, that is applicable for any operating system.
The intuitive Graphical User Interface (GUI) of \name is another metric of the user experience. The overall average score of 8.35 for this metric shows that the students has had no difficulty in dealing with the \name through its GUI. As per gender assessment, female students assessed the GUI intuitive and easy to use with the average score of 9.3 while male students rated it as 8. Moreover, the average score of 8.3 (female average score:9.3, male average score: 7.9) for ease-of-use metric demonstrate that students assess the overall technical part of E2C is intuitive and easy to understand. However, the students assessed the report section with the average score of 5.7 (female average score:4.8, male average score: 5.9). Although the reports are comprehensive, we realized that the structure of the GUI for the report section is not intuitive, therefore, the students could not find their required reports easily. To address this issue, we are rearranging the report section in the GUI and make different reports and their fields more informative. In case of developing a custom scheduling in E2C, the graduate students responded that \name was useful, with the average score of 8.3 (female average score:9.2, male average score: 7.4), in implementing and evaluating their custom scheduling policy. In general, the students evaluated their willingness for recommending \name to others with the average score of 8.3 (female average score: 9.7, male average score: 7.8), as shown in Figure~\ref{fig:hci}.

Figure~\ref{fig:learning} summarizes the students' responses in terms of their learning outcomes. The results show that they found \name helpful in understanding the impact of scheduling methods in heterogeneous and homogeneous systems with the median score of 8.7 (female average score:9.8, male average score: 8.2) and 8 (female average score:9.5, male average score: 8.4), respectively. In addition, as explained in Section~\ref{sec:assign}, they utilized three workload traces with varying arrival intensities to learn about the impact of arrival rate on the system performance in terms of on-time completion rate. As shown in Figure~\ref{fig:learning}, they responded that \name could help them in understanding the impact of arrival rate on the system performance with the average score of 8.6 (female average score:9.7, male average score: 8.2). Overall, based on the survey results, shown in Figure~\ref{fig:learning}, students assessed E2C is useful in developing their knowledge in the distributed systems course with the median score of 8.8 (female average score:9.5, male average score: 8.6). More specifically, as shown in the results, female students assessed \name as a an easy-to-use and useful learning tool with higher median score than male students. In other words, the gender-based results show that \name is more effective for female students. 

We asked students similar scheduling questions in the form of two quizzes, taken before and after using \name as a course assignment. The quizzes asked the students to map three arriving tasks to four heterogeneous machines via the following scheduling methods: MEET, MECT, MM, and MSD. The average score of students has improved from 7.6 (out 12 points) in the first quiz to 8.94 in the second quiz. The results imply that \name could improve the students' learning of scheduling methods in heterogeneous computing systems by 17.6\%.

At the end of the survey study, we asked them to write us their feelings and suggestions that they would like to see in the next version of the \name simulator. Here, is the main suggestions we received from them:
``The simulator clarified the working of different scheduling methods well with its visual animation."
``The application was intuitive when it comes to the context of this course and it was relatively easy to use." ``I must commend the great work done by the everyone at the HPCC lab that contributed to
the E2C simulator. This is a wonderful software."
As for the suggestions, students reported several bugs that we already fixed. Some others had suggestions to make the GUI more intuitive, \eg by changing the mouse pointer when it is hovered on various components; also there were suggestions to enable drag and drop feature to the simulation scenario.

\section{\name Code and Resource Availability}\label{sec:code}
\name core is available for download at the following address:
\smallskip\noindent
\footnotesize \texttt{\url{https://github.com/hpcclab/E2C-Sim}}
\normalsize

\smallskip \noindent
The manual document on how to run \name and its options and full documentations are available here:

\smallskip\noindent
\footnotesize \texttt{\url{https://hpcclab.github.io/E2C-Sim-docs/}}
\normalsize

\smallskip \noindent
The video resources for \name are in this  \href{https://youtube.com/playlist?list=PL7jhdCPVrCHh49PvIglDEY2Xs4v2ivrsw}{YouTube page}.

\section{Conclusion and Future Works}\label{sec:conc}
\name provides a free (open-source) learning tool for students enrolled in courses like Distributed Systems, Operating Systems, and Computer Networks as well as researchers by delivering an intuitive way to simulate heterogeneous and homogeneous systems. It particularly helps the students to gain insight into the performance of different scheduling methods upon various heterogeneous systems and under various workload intensities without the need to use and expend for real infrastructure. As such \name is a step towards reducing the widening educational gap nationally, and even at the global scale.
The users of this system can employ several existing scheduling methods built into the simulator, but also have the ability to develop and test their own custom method. As we experienced it in our Distributed and Cloud Computing class, it is an effective accompaniment that can remarkably improve the knowledge of students in the area of heterogeneous computing and scheduling. \name comes with user friendly GUI for quick usage by beginners, but is also configurable enough to meet the needs of researchers and practitioners in the field.
Based on the feedback we received from our students, we plan to extend \name with several other features, including various communication paradigms and the ability to drag and drop components into the simulator.

\section*{acknowledgement}
Development of \name was made possible by the funding support provided by National Science Foundation (NSF) under awards\# CNS-2007209 and CNS-2047144 (NSF CAREER Award).

\bibliographystyle{plain} 
\bibliography{reference}

\end{document}